# Project Blue – Visible Light Imaging Search for Terrestrial-class Exoplanets in the Habitable Zones of Alpha Centauri A & B

Jon A. Morse (BoldlyGo Institute, Project Blue Mission Executive)
Ph: 646-380-1813, Email: jamorse@boldlygo.org

Co-authors:

    Eduardo Bendek (Bay Area Environmental Research Institute)
    Nathalie Cabrol, Franck Marchis, Margaret Turnbull (SETI Institute)
    Supriya Chakrabarti (UMass Lowell)
    Debra Fischer (Yale University)
    Colin Goldblatt (University of Victoria)
    Olivier Guyon, Michael Hart, Jared Males (University of Arizona)
    Jim Kasting (Penn State University)

# Project Blue – Visible Light Imaging Search for Terrestrial-class Exoplanets in the Habitable Zones of Alpha Centauri A & B

## A QUEST TO SEE BLUE

*Finding the first planet like Earth beyond our Solar System will be an historic scientific achievement and will transform how we think about our place in the universe. Finding such a planet in the Alpha Centauri system would lay the foundation and motivation for humanity's next millennium of deep space exploration.*

Project Blue brings together a consortium of space research organizations and scientists who will work with NASA, aerospace companies and international partners to design, build, launch and operate a small, state-of-the-art space telescope to search for exoplanets around our nearest stellar neighbors: Alpha Centauri A (stellar type G2 V; $V_{mag}$=+0.01; also known as Rigil Kentaurus) and B (K1 V; $V_{mag}$=+1.33). Project Blue's goal is to perform a direct imaging survey of the habitable zones of Alpha Cen A & B in visible light, to determine whether a "pale blue dot" exists around our nearest stellar neighbors. At modest cost compared to other space missions and with a potential launch in the early 2020's, this goal is tantalizingly close.

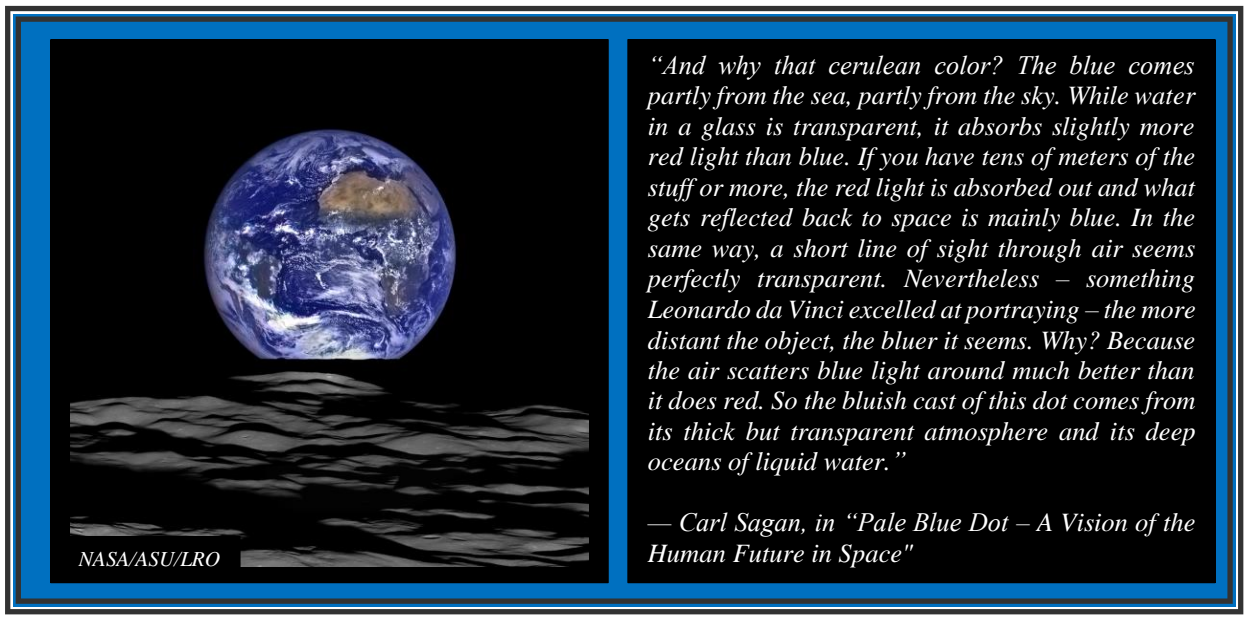

NASA/ASU/LRO

*"And why that cerulean color? The blue comes partly from the sea, partly from the sky. While water in a glass is transparent, it absorbs slightly more red light than blue. If you have tens of meters of the stuff or more, the red light is absorbed out and what gets reflected back to space is mainly blue. In the same way, a short line of sight through air seems perfectly transparent. Nevertheless – something Leonardo da Vinci excelled at portraying – the more distant the object, the bluer it seems. Why? Because the air scatters blue light around much better than it does red. So the bluish cast of this dot comes from its thick but transparent atmosphere and its deep oceans of liquid water."*

*— Carl Sagan, in "Pale Blue Dot – A Vision of the Human Future in Space"*

## SCIENCE & OBSERVATORY DESCRIPTION

### COMPELLING SCIENCE, PATHFINDING TECHNOLOGY

Project Blue will advance our knowledge about the presence or absence of terrestrial-class exoplanets in the habitable zones around the nearest Sunlike stars in the Alpha Cen system (d = 1.34 pc; see Figure 1). We aim to image terrestrial-class planets (e.g., in the range 0.5 to 2 Earth-radii), if planets remain in the system at all, since kinematic measurements exclude habitable zone planets more massive than ~8 $M_\oplus$ for Alpha Cen B and ~50 $M_\oplus$ for A (Zhao et al. 2017). The mission also will provide on-orbit demonstration of high-contrast coronagraphic imaging technologies and techniques that will be useful for planning future space missions by NASA and other space agencies. The mission concept overlaps with the *ACESat* concept described in Belikov et al. (2015) and Bendek et al. (2015), which we recommend for further reading.

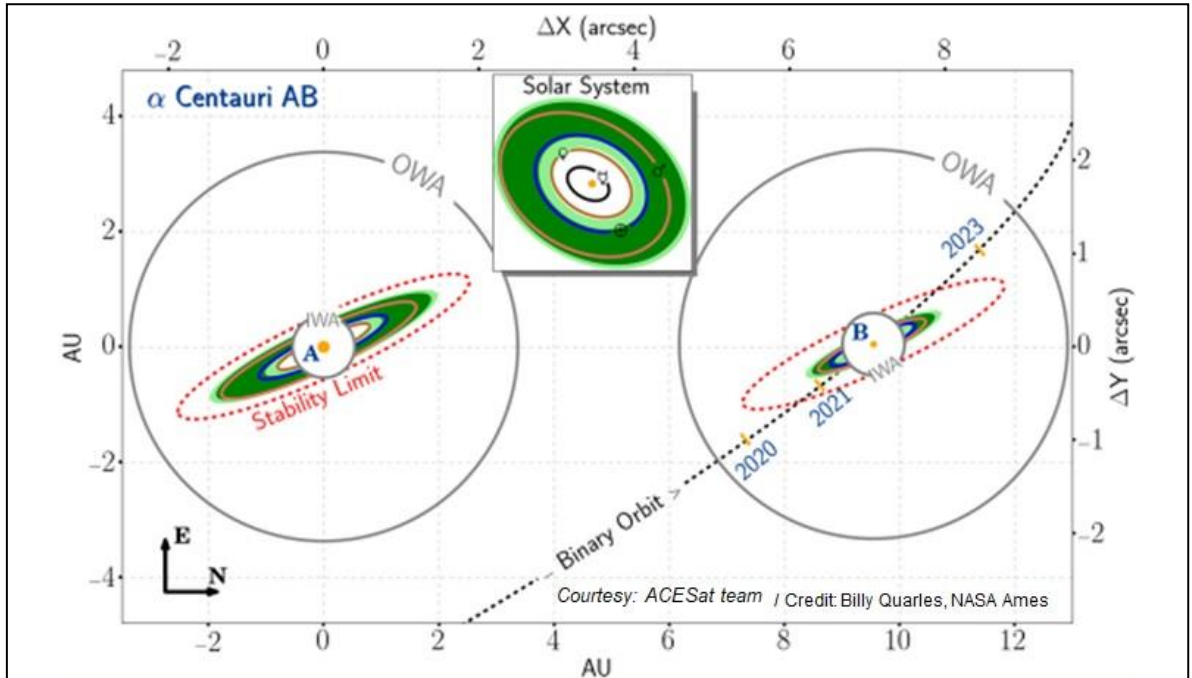

*Figure 1. Illustration of the regions that Project Blue will search for terrestrial-class planets around the two Sunlike stars in the Alpha Centauri system. The habitable zones where liquid water could exist on the surface of a terrestrial sized planet are shown as the green rings around Alpha Cen A (left) and B (right), with our Solar System (top-center) shown for comparison. The dashed oval is the Stability Limit, inside which planets and zodiacal dust could survive on Gyr timescales (e.g., Quarles & Lissauer 2016).*

The Project Blue investigation aligns with the exoplanet science priorities described in the *New Worlds New Horizons* decadal survey as well as recent reports by NASA. These reports stake out a path of increasingly more ambitious measurements, from identifying candidate Earthlike planets to assessing their atmospheres for biomarkers. Project Blue seeks to make the initial historic leap of identifying Earthlike planet candidates in the Alpha Cen system. The observations can yield motions on the plane of the sky along with the brightness and visual color of planet candidates. These measurements can be used to infer planetary characteristics such as orbit, size, albedo, surface & atmospheric properties, and possibly mass. The observations will also tell us about the environment in which any such planets orbit. Besides searching for exoplanets around Alpha Cen A & B, Project Blue could directly measure the brightness of the zodiacal dust around each star, which will aid future missions in planning their observational surveys.

## OBSERVATORY DESCRIPTION

In order to spatially resolve the habitable zones around Alpha Cen A & B, the Project Blue mission needs to fly a space telescope only 45-50 cm in aperture, small enough to fit on a coffee table. As our name suggests, observations will be made in blue light ($\lambda_{cen}$~465nm) and one or two longer wavelength bands to capture the hue of any planets discovered.

*"NASEM/SSB Exoplanet Science Strategy"*
*Project Blue Overview – March 2018*

An overview the Project Blue mission concept is shown in Figure 2. The telescope is an off-axis Ritchey-Chrétien design optimized for coronagraphic observations, serving as a pathfinder for larger off-axis telescope designs such as HabEx or Exo-C. Our 2017 concept study verified that such an observatory, using a commercially available spacecraft bus, could fit within the volume and lift capacity of a one or more launch vehicles expected to be available in the early 2020's.

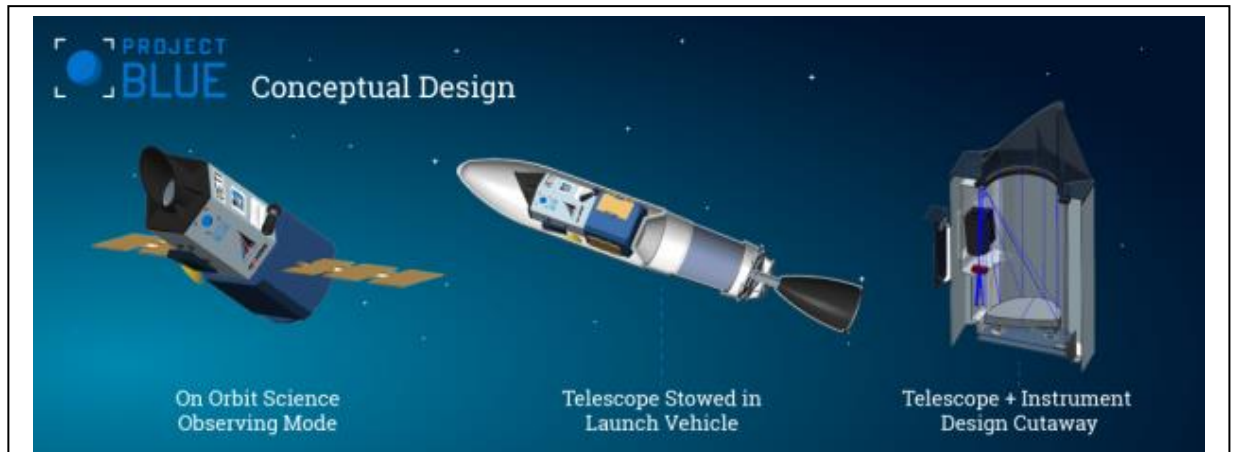

*Figure 2. Different aspects of the Project Blue space observatory, showing its deployed on-orbit configuration for science observing (left), stowed configuration inside the fairing attached to the upper stage of a representative small launch vehicle (center), and cutaway schematic of the path that light rays travel through the telescope to be recorded on a digital coronagraphic camera (right).*

Our baseline is to use a Phase Induced Amplitude Apodization (PIAA; e.g., Guyon et al. 2006) design, to complement the WFIRST coronagraph approach, with a starlight suppression system consisting of (1) a deformable mirror (DM) that modifies the wavefront of the incoming light, (2) a coronagraph to block the parent starlight, and (3) an image processing methodology, called orbital differential imaging (ODI; Males et al. 2015), incorporating aspects of AI pattern recognition to detect faint exoplanets against background sources in the vicinity of the parent star.

These components of the imaging system, data acquisition and post-processing work in tandem to achieve our planet imaging goals. We require the coronagraph to achieve contrasts of $\sim 10^{-8}$ in the raw data, and believe that post-processing can wring an additional $10^{-2}$ or more out of the data (see Figure 3). An essential aspect of our approach is that we plan to observe Alpha Cen for two full years, obtaining a vast database of imagery that will allow scientists to retrieve the faint signal of any planets by beating down the residual noise from the optical system, and to see the planets move in proper motion with their parent star as well as in their orbits. The habitable zones of Alpha Cen A & B can be searched over an angular range of 0.6″ – 1.5″ (from $3\lambda/D$ to $\sim 8\lambda/D$) corresponding to $\sim 0.8 – 2$ AU. An inner working angle of 0.4″ (0.53 AU) could be probed if the PIAA coronagraph performs down to $2\lambda/D$.

The Project Blue mission is technically challenging, but all aspects of the observatory have considerable heritage, are in family with previous missions and instrumentation, and/or have significant recent technological maturity investments. Project Blue team members have significant experience developing and using coronagraphic imaging technologies in laboratory experiments, ground-based telescopes, and even with suborbital sounding rocket and balloon payloads. Flight experiments have used deformable mirrors, demonstrated precision pointing control and nulling interferometer coronagraphy, and developed complex post-processing of the data. Project Blue will need to suppress the light of both Alpha Cen A & B simultaneously; one way to attack this challenge is through multi-star wavefront control (MWSC; Sirbu et al. 2017), an algorithm under development with NASA funding to support coronagraphic observations of multi-star systems with WFIRST, HabEx and LUVOIR.

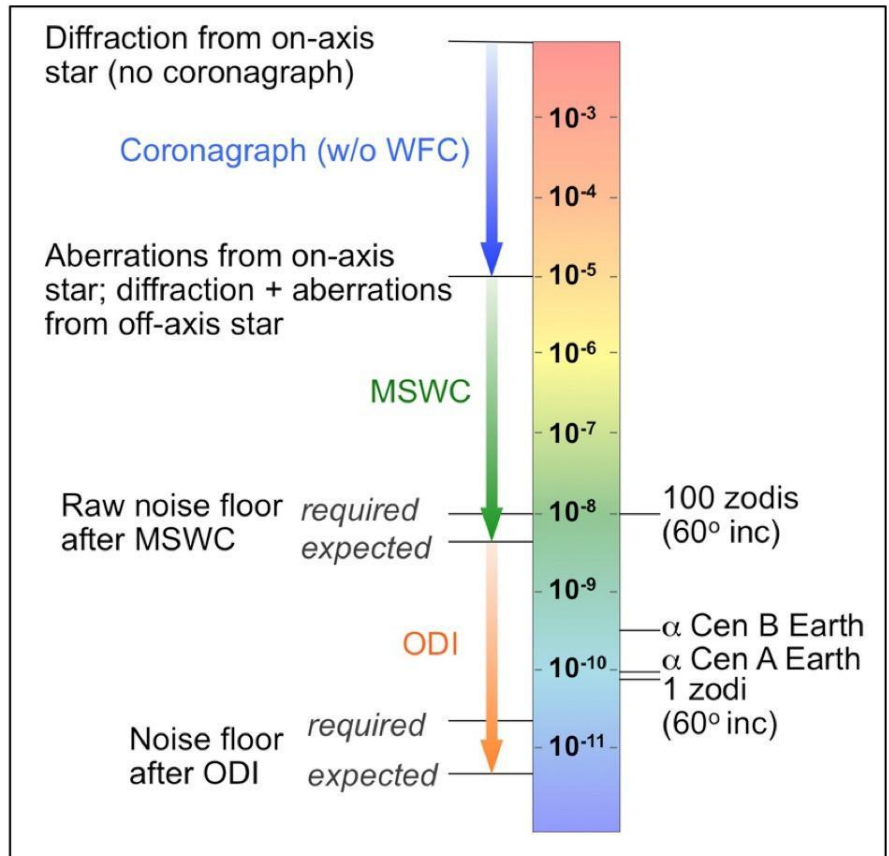

*Figure 3. Starlight suppression contributions of optics, deformable mirrors with multi-star wavefront control (MSWC) and orbital differential imaging (ODI) post-processing. (Courtesy ACESat team; Bendek et al 2015).*

Like NASA missions, we plan to make the raw and processed data publically available through an online archive that will allow scientists and citizens to engage in their own analyses. This is one of the provisions in our nonreimbursable Space Act Agreement with NASA.

Because the telescope is customized for intensive observing of Alpha Centauri, certain design and operational aspects of the mission can be simplified compared to larger missions conducting surveys of dozens of exoplanetary systems along with other astrophysical investigations. Focusing on one target allows us to design specific system calibration and noise-reduction algorithms. In addition, the modest telescope size and our plan to leverage substantial commercial space capabilities will allow us to lower costs and shorten the development schedule relative to much larger missions. Our goal is a 3-4 yr development time within a NASA Mission of Opportunity budget envelope (~$70M).

## PROGRAMMATIC CONSIDERATIONS

Alpha Centauri presents a compelling opportunity to search for terrestrial-class exoplanets in the habitable zones of the two nearest Sunlike stars. If humanity eventually goes to the stars – first with robots, and perhaps in the far-future with crewed spacecraft – the presence (or absence) of a destination in the Alpha Cen system will play an important, even decisive, role. We note that such a null result, though typically not an important factor in scientific review panels, may be extremely important to humanity's very long-term exploration goals.

While we are considering traditional funding paths and pursuing international partnerships, we believe that the potential legacy of Project Blue is great enough that we are actively fundraising private philanthropic resources to support the Project Blue mission development and science investigation. Our aim is to enable a public-private partnership that accelerates the pace of exoplanet discovery and reduces the budget burden on federal resources. Like many science perspectives, technology initiatives, and mission concepts submitted to the committee for consideration, our project will benefit from close coordination and active collaboration with NASA scientists and engineers.

But are NASA and the community willing to take certain potentially high-payoff risks, or consider alternative funding mechanisms such as funded Space Act Agreements or data buys? (Our experience even suggests that a suitable cash prize could effectively attract venture capital.) And are we collectively willing to fund certain pathfinder missions (e.g., ESA's LISA Pathfinder) whose main impact may be to demonstrate technologies, reduce risk and hopefully lower costs of future large missions? This may be the price of achieving a truly historic goal.

For more information about the Project Blue coronagraphic space telescope mission and team, please visit http://www.projectblue.org.